\documentclass[12pt]{article}
\usepackage{mathrsfs}
\usepackage{amsfonts}
\usepackage{amsmath}
\usepackage{amssymb}
\usepackage{mathptmx}
\usepackage[dvips]{color}
\usepackage{epsfig}
\usepackage{graphicx}
\newcommand{\calA}{{\cal A}}
\newcommand{\calF}{{\cal F}}
\newcommand{\calG}{{\cal G}}
\newcommand{\calL}{{\cal L}}

\newcommand{\kvec}{{\bf k}}

\newcommand{\GeV}{{\rm GeV}}

\begin{document}
\baselineskip=16pt

\pagenumbering{arabic}

\vspace{1.0cm}

\begin{center}
{\Large\sf Multi-photon decays of the Higgs boson in standard model:
leading terms from Heisenberg-Euler effective Lagrangian}
\\[10pt]
\vspace{.5 cm}

{Yi Liao\footnote{liaoy@nankai.edu.cn}}

{\it School of Physics, Nankai University, Tianjin 300071, China}

\vspace{2.0ex}

{\bf Abstract}

\end{center}

We calculate the multi-photon decay widths of the Higgs boson from
an effective Lagrangian for a system of electromagnetic and Higgs
fields. We utilize a low-energy theorem to connect the above
Lagrangian to the Heisenberg-Euler effective Lagrangian induced by
charged particles that gain mass from interactions with the Higgs
boson. In the standard model only the $W^\pm$ gauge bosons and the
top quark are relevant, and we compute their contributions to the
effective couplings and the multi-photon decay widths of the Higgs
boson.

\begin{flushleft}

\end{flushleft}

\newpage

A new boson has recently been discovered at the CERN Large Hadron
Collider (LHC) \cite{:2012gk,:2012gu}, whose properties are found to
be close to those expected for the Higgs boson, $h$, in the standard
model (SM) \cite{Weinberg:1967tq,Glashow:1961tr,Salam:1968rm}. The
rare two-photon decay of the Higgs boson has offered a golden
detection channel in this great endeavor; for a review on Higgs
phenomenology see for instance \cite{Gunion:1989we}. As more events
are being accumulated, its detailed properties will be uncovered.
More optimistically, a linear collider or even a dedicated Higgs
factory will be very helpful to reveal whether the particle is
genuinely the last missing piece of SM or the first new member in a
bigger picture. For this purpose, it is necessary to examine
exhaustively its production and decay properties. In this work we
study its multi-photon decays which are closely related to the
two-photon channel.

As a neutral particle, the photonic decay of the Higgs boson is a
purely quantum effect. In SM the decay arises at the one-loop level
through interactions with the vacuum charged particles which gain
mass through the Higgs mechanism
\cite{Higgs:1964pj,Englert:1964et,Guralnik:1964eu}. This last
feature implies that the heaviest charged particles, the top quark
and the $W^\pm$ gauge bosons, dominate the decay process. The
two-photon decay rate in SM was computed long time ago
\cite{Ellis:1975ap,Ioffe:1976sd,Shifman:1979eb,Vainshtein:1980ea,Rizzo:1979mf}.
As the number of photons increases, the computation rapidly becomes
technically more and more challenging. This arises due mainly to the
constraint from electromagnetic gauge invariance, which results in
significant cancellation amongst the contributions from individual
Feynman graphs \cite{Karplus:1950zz}. Fortunately, it has also been
known for long \cite{Ellis:1975ap,Shifman:1979eb} that the
two-photon decay amplitude can be reasonably well approximated by
the leading terms in the expansion of the Higgs mass $m_h$ over
twice masses of the relevant heavy charged particles, i.e.,
$2m_{t,W}$; for instance, for $m_h=125~\GeV$ as indicated by the
measurements at LHC, the leading terms account for about $65\%$ of
the exact one-loop decay width. With more photons in the final state
sharing the Higgs boson mass and thus becoming softer on average, we
expect that the expansion in $r_{t,W}=[m_h/(2m_{t,W})]^2$ works even
better. In this work we report the results for the four- and
six-photon decay rates in this approximation.

The low-energy effective Lagrangian for the system of the Higgs
boson and electromagnetic fields starts with the terms that involve
the least number of derivatives for a given number of fields while
respecting electromagnetic gauge invariance. Since the Higgs field
$h$ is neutral, the electromagnetic field must form into gauge
invariants in itself. There are two such basic invariants involving
the least number of derivatives,
$\calF=\frac{1}{4}F_{\mu\nu}F^{\mu\nu}$ and
$\calG=\frac{1}{8}\epsilon^{\mu\nu\alpha\beta}F_{\mu\nu}F_{\alpha\beta}$,
which are respectively a Lorentz scalar and pseudoscalar. Attaching
derivatives to $h$, $\calF$ or $\calG$ does not harm the gauge
symmetry but will yield terms of a higher order in the low-energy
expansion. We assume that charge conjugation (C) and parity (P) are
good symmetries, which is the case in SM at the level under
consideration. Then, the Higgs boson as a C- and P-even particle can
only decay into an even number of photons, and the pseudoscalar
invariant $\calG$ must appear as an even power in interactions with
the Higgs boson. The leading terms in the single-$h$ sector can
therefore be parameterized as follows,
\begin{eqnarray}
\calL_h&=&\xi_{10}h\calF%
+h\Big[\frac{1}{2!}\xi_{20}\calF^2+\frac{1}{2!}\xi_{02}\calG^2\Big]
+h\Big[\frac{1}{3!}\xi_{30}\calF^3+\frac{1}{2!}\xi_{12}\calF\calG^2\Big]
\nonumber\\
&&+h\Big[\frac{1}{4!}\xi_{40}\calF^4+\frac{1}{2!2!}\xi_{22}\calF^2\calG^2
+\frac{1}{4!}\xi_{04}\calG^4\Big]+\cdots
\label{eq_Lagh}
\end{eqnarray}
The effective coupling $\xi_{ij}$ accompanies the form
$\calF^i\calG^j$, and is proportional to
$e^{2(i+j)}y_hM^{3-4(i+j)}$, where $e$ is the electromagnetic
coupling, $y_h$ is the Higgs coupling to the relevant heavy charged
fields (top quark and $W^\pm$ bosons in SM) of mass $M$, and a sum over
them is implied.

To present the decay amplitudes in a compact form, we introduce the
notations for a pair of photons, $ij$, with outgoing momenta
$k_i,~k_j$ and polarization vectors $\epsilon_i,~\epsilon_j$,
\begin{eqnarray}
E_{ij}&=&k_i\cdot\epsilon_j^*k_j\cdot\epsilon_i^* -k_i\cdot
k_j\epsilon_i^*\cdot\epsilon_j^*,
\\
O_{ij}&=&\epsilon_{\mu\nu\alpha\beta}k_i^\mu k_j^\nu
\epsilon_i^{*\alpha}\epsilon_j^{*\beta}
\end{eqnarray}
A factor of $\calF$ ($\calG$) in $\calL_h$ when assigned to the
photon pair $ij$ will yield in the amplitude a factor of $E_{ij}$
($O_{ij}$). Note that both notations are symmetric under the
interchange of the two photons. The amplitude for the two-photon
decay $h\to 12$ is
\begin{eqnarray}
\calA_{12}&=&\xi_{10}E_{12}
\end{eqnarray}
The results for the decay into more photon pairs are similar with
the only difference being in combinatorics. For instance, for the
four-photon decay $h\to 1234$,
\begin{eqnarray}
\calA_{1234}&=&\xi_{20}E_{1234}+\xi_{02}O_{1234},
\end{eqnarray}
where
\begin{eqnarray}
E_{1234}&=&E_{12}E_{34}+E_{13}E_{24}+E_{14}E_{23},
\\
O_{1234}&=&O_{12}O_{34}+O_{13}O_{24}+O_{14}O_{23},
\end{eqnarray}
and for the six-photon decay $h\to 123456$,
\begin{eqnarray}
\calA_{12\cdots 6}&=&\xi_{30}E_{12\cdots 6}
+\xi_{12}\big[E_{12}O_{3456}+\textrm{14 perms.}\big],
\end{eqnarray}
where
\begin{eqnarray}
E_{12\cdots 6}&=&
E_{12}\big(E_{34}E_{56}+E_{35}E_{46}+E_{36}E_{45}\big) +\textrm{4
perms.}
\end{eqnarray}

The sum over photons' polarizations in the amplitudes squared is
complicated, and has been done with the help of a computer code. We
find for the two-photon decay,
\begin{eqnarray}
\sum|\calA_{12}|^2&=&\xi_{10}^22U(12),
\end{eqnarray}
where
\begin{eqnarray}
U(12)&=&(12)^2
\end{eqnarray}
with $(ij)\equiv k_i\cdot k_j$. The pure-$E$ and pure-$O$ terms for
the  four-photon decay are found to contribute the same, so that
\begin{eqnarray}
\sum|\calA_{1234}|^2&=&\big[\xi_{20}^2+\xi_{02}^2\big]\big[10U(1234)-8X(1234)\big]
\nonumber\\
&&+2\xi_{20}\xi_{02}\big[-6U(1234) +8X(1234)\big],
\end{eqnarray}
where
\begin{eqnarray}
U(1234)&=&\sum_\textrm{3 perms.}(ij)^2(mn)^2
=(12)^2(34)^2+\textrm{2 perms.},
\\
X(1234)&=&\sum_\textrm{3 perms.}(ij)(jm)(mn)(ni)
=(12)(23)(34)(41)+\textrm{2 perms.}
\end{eqnarray}
The result for the six-photon case is fairly lengthy:
\begin{eqnarray}
&&\sum|\calA_{1\cdots 6}|^2
\nonumber\\
&=&\xi_{30}^2\big[44U(1\cdots 6)
+24V(1\cdots 6)+120Y(1\cdots 6)-56X(1\cdots 6)\big]
\nonumber\\
&&+\xi_{12}^2\big[348U(1\cdots 6)-8V(1\cdots 6)+504Y(1\cdots 6) -216X(1\cdots 6)\big]
\nonumber\\
&&+2\xi_{30}\xi_{12}\big[-36U(1\cdots 6)-104V(1\cdots 6)-360Y(1\cdots 6)+168X(1\cdots 6)\big],
\end{eqnarray}
where the functions of six-photons' momenta are
\begin{eqnarray}
U(1\cdots 6)&=&\sum_\textrm{15 perms.}(ij)^2(mn)^2(pq)^2,
\\
V(1\cdots 6)&=&\sum_\textrm{15 perms.}U(ij)X(mnpq),
\\
X(1\cdots 6)&=&\sum_\textrm{60 perms.}(ij)(jm)(mn)(np)(pq)(qi),
\\
Y(1\cdots 6)&=&\sum_\textrm{10 perms.}(ij)(jp)(pi)(mn)(nq)(qm)
\end{eqnarray}

The phase-space integral for an all-massless final state can be
completely finished for a polynomial of momenta. An elegant method
was provided in \cite{Aghababaie:2000zi}. Denoting the $n$-photon
phase-space integral as
\begin{eqnarray}
\textrm{PS}_n=\frac{1}{n!}\int\prod_{j=1}^n\bigg[\frac{d^3\kvec_j}{(2\pi)^32|\kvec_j|}\bigg]
(2\pi)^4\delta^4\Big(p-\sum_{i=1}^nk_i\Big),
\end{eqnarray}
where $p$ is the incoming momentum of the Higgs boson, the basic
idea is to decouple the correlation amongst the photons' momenta  by
converting the delta function back to an integral over the
coordinate $x$ and introducing a separate integral for each photon
\cite{Aghababaie:2000zi},
\begin{eqnarray}
J(x)=\int\frac{d^3k}{2|\kvec|}e^{-ik\cdot x}=-\frac{2\pi}{x^2},
\end{eqnarray}
where an infinitesimal, negative imaginary part is attached to $x_0$
to make the integral well-defined. The factors of $k$ are induced by
derivatives with respect to $x$, for instance, for our study here,
the following will be required,
\begin{eqnarray}
J^{\alpha\beta}(x)=\int\frac{d^3k}{2|\kvec|}k^\alpha k^\beta
e^{-ik\cdot x}=i^2\partial^\alpha\partial^\beta J(x),
\end{eqnarray}
and the scalar products of different photons' momenta in the
phase-space integral are obtained by appropriate contractions with
the signature tensor. The final  central integral is the following
one for $m\ge 2$ \cite{Aghababaie:2000zi}:
\begin{eqnarray}
I_m(p)=\int d^4x\frac{e^{ip\cdot x}}{(x^2)^m}
=\frac{(-1)^m 2^{5-2m}\pi^3}{(m-1)!(m-2)!}(p^2)^{m-2}
\end{eqnarray}

We list the integrals that are required for the Higgs decays up to six photons:
\begin{eqnarray}
\big[2\gamma\big]_1&=&\textrm{PS}_2(12)^2
=\frac{(4\pi)^2}{2!(2\pi)^6}~12I_4(p),
\\
\big[4\gamma\big]_1&=&\textrm{PS}_4(12)^2(34)^2
=\frac{(4\pi)^4}{4!(2\pi)^{12}}~144I_8(p),
\\
\big[4\gamma\big]_2&=&\textrm{PS}_4(12)(23)(34)(41)
=\frac{(4\pi)^4}{4!(2\pi)^{12}}~84I_8(p),
\\
\big[6\gamma\big]_1&=&\textrm{PS}_6(12)^2(34)^2(56)^2
=\frac{(4\pi)^6}{6!(2\pi)^{18}}~1728I_{12}(p),
\\
\big[6\gamma\big]_2&=&\textrm{PS}_6(12)(23)(34)(45)(56)(61)
=\frac{(4\pi)^6}{6!(2\pi)^{18}}~732I_{12}(p),
\\
\big[6\gamma\big]_3&=&\textrm{PS}_6(12)^2(34)(45)(56)(63)
=\frac{(4\pi)^6}{6!(2\pi)^{18}}~1008I_{12}(p),
\\
\big[6\gamma\big]_4&=&\textrm{PS}_6(12)(23)(31)(45)(56)(64)
=\frac{(4\pi)^6}{6!(2\pi)^{18}}~576I_{12}(p)
\end{eqnarray}
Noting that the terms obtained by permutation of the photons'
momenta contribute  the same in the phase-space integrals, the decay
widths are worked out to be
\begin{eqnarray}
\Gamma(2\gamma)&=&\frac{m_h^3}{2^6\pi}\xi_{10}^2,
\\
\Gamma(4\gamma)&=&\frac{m_h^{11}}{2^{20}~3^3~5^2~7\pi^5}
\big[2\big(\xi_{20}^2+\xi_{02}^2\big)-\xi_{20}\xi_{02}\big],
\\
\Gamma(6\gamma)&=&\frac{m_h^{19}}{2^{36}~3^8~5^3~7^2~11\pi^9}\big[2\xi_{30}^2
+9\xi_{12}^2-3\xi_{30}\xi_{12}\big]
\end{eqnarray}

The above results apply to any theory that induces effective
interactions  between a single Higgs scalar and various numbers of
photons that preserve C and P symmetries as shown in eq.
(\ref{eq_Lagh}). In the following, we will show the numerical
results in SM. To this end, we have to work out first the $\xi$
parameters. Since the fermions and weak gauge bosons in SM gain mass
completely through interactions with the Higgs field which develops
a vacuum expectation value, $v$, their mass terms and interactions
with the Higgs boson appear in the combined form $m(1+h/v)$ to a
linear or quadratic power. This implies a low-energy theorem
\cite{Shifman:1979eb} between a process involving a Higgs particle
and one without. For a system of constant electromagnetic field
strength and Higgs field, we can thus establish a relation between
the effective Lagrangian $\calL_h$ in eq. (\ref{eq_Lagh}) and the
Heisenberg-Euler effective Lagrangian $\calL_\textrm{H.-E.}$:
\begin{eqnarray}
\calL_h=\frac{h}{v}\bigg(\sum_fm_f\frac{\partial}{\partial m_f}
+\sum_Vm_V\frac{\partial}{\partial m_V}\bigg)\calL_\textrm{H.-E.}
=\frac{g_2h}{m_W}\bigg(\sum_fm_f^2\frac{\partial}{\partial m_f^2}
+\sum_Vm_V^2\frac{\partial}{\partial m_V^2}\bigg)\calL_\textrm{H.-E.}%
\label{eq_let}
\end{eqnarray}
The mass-proportionality means that only the top quark and $W^\pm$
bosons are actually relevant; namely, for photonic Higgs decays, we
only  have to include in $\calL_\textrm{H.-E.}$ the polarization
effects due to vacuum top quarks and $W^\pm$ bosons:
\begin{eqnarray}
\calL_\textrm{H.-E.}=\calL_\textrm{H.-E.}^t+\calL_\textrm{H.-E.}^{W^\pm}+\cdots
\end{eqnarray}

The vacuum polarization effects due to top quarks
$\calL_\textrm{H.-E.}^t$ can be readily obtained from the QED result
\cite{Heisenberg:1935qt,Schwinger:1951nm}, see \cite{Dunne:2004nc}
for a review, by the substitutions of the electron parameters with
the top ones, $m_e\to m_t$ and $-e\to Q_te$ with $Q_t=2/3$, and
attaching a color factor $N_c=3$:
\begin{eqnarray}
\calL^t_\textrm{H.-E.}&=&-\frac{N_c}{8\pi^2}\int_0^\infty\frac{ds}{s}e^{-m_t^2s}
\frac{Q_t^2e^2F_-F_+}{\tan(Q_teF_-s)\tanh(Q_teF_+s)},
\end{eqnarray}
where
\begin{eqnarray}
F_\pm=\sqrt{\sqrt{\calF^2+\calG^2}\pm\calF}
\end{eqnarray}
The vacuum polarization effects due to a charged vector field were
first studied in \cite{Vanyashin} for a general electromagnetic
theory of a massive charged vector field. It was found that the
problem was exactly solvable only when the charged vector has a
gyromagnetic ratio equal to 2 whence the charge renormalization
constant was found to be larger than one in contrast to QED. In
retrospect, the first feature means that a consistent theory of
charged vector fields has to be a non-Abelian gauge theory while the
second implies asymptotic freedom in such a theory. The issue was
later reexamined in a spontaneously broken gauge theory in unitary
gauge \cite{Skalozub:1975ab} and nonlinear $R_\xi$ gauge
\cite{Skalozub:1980nv}, see \cite{Skalozub:1986he} for a review on
the topic and related ones. The effective Lagrangian was found to be
\begin{eqnarray}
\calL_\textrm{H.-E.}^{W^\pm}&=&-\frac{1}{16\pi^2}\int_0^\infty\frac{ds}{s}e^{-m_W^2s}
\frac{e^2F_-F_+\big[1-2\cos(2eF_-s)-2\cosh(2eF_+s)\big]}{\sin(eF_-s)\sinh(eF_+s)}
\end{eqnarray}

Upon expansion in the electromagnetic field, the low-energy theorem
in eq. (\ref{eq_let}) then yields the dominant contributions in SM
to the couplings in eq. (\ref{eq_Lagh}):
\begin{eqnarray}
\xi_{10}&=&\frac{g_2}{m_W}\frac{e^2}{(4\pi)^2}\Big[-7+\frac{4}{3}Q_t^2N_c\Big],
\\
\xi_{20}&=&\frac{g_2}{m_W^5}\frac{2e^4}{(4\pi)^2}\Big[-\frac{29}{5}
-\frac{16}{45}Q_t^4N_c\frac{m_W^4}{m_t^4}\Big],
\\
\xi_{02}&=&\frac{g_2}{m_W^5}\frac{2e^4}{(4\pi)^2}\Big[-\frac{27}{5}
-\frac{28}{45}Q_t^4N_c\frac{m_W^4}{m_t^4}\Big],
\\
\xi_{30}&=&\frac{g_2}{m_W^9}\frac{6e^6}{(4\pi)^2}\Big[-\frac{548}{105}
+\frac{256}{315}Q_t^6N_c\frac{m_W^8}{m_t^8}\Big],
\\
\xi_{12}&=&\frac{g_2}{m_W^9}\frac{2e^6}{(4\pi)^2}\Big[-\frac{628}{105}%
+\frac{416}{315}Q_t^6N_c\frac{m_W^8}{m_t^8}\Big]
\end{eqnarray}
Two features are noteworthy. First, it seems that the $W^\pm$
contribution to the effective couplings is always negative while the
top-quark contribution is alternate in sign as the power of the
electromagnetic invariants $\calF,~\calG$ increases. This is in
contrast with the impression that one might have got from the
destructive interference between the fermionic and bosonic loops in
the two-photon decay channel. Second, because of larger mass and
smaller charge the top-quark contribution becomes more and more
suppressed as the power of invariants increases. For instance, its
relative contribution in the cubic terms already drops down to
$10^{-4}$ and can be totally ignored. At $m_h=125~\GeV$, these
parameters yield the following approximate photonic decay widths of
the Higgs boson in SM:
\begin{eqnarray}
\Gamma(2\gamma)&=&5.9\times 10^{-6}~\GeV
\\
\Gamma(4\gamma)&=&3.1\times 10^{-15}~\GeV
\\
\Gamma(6\gamma)&=&8.5\times 10^{-27}~\GeV
\end{eqnarray}

We have studied the multi-photon decays of the Higgs boson that are
closely related to the two-photon decay employed as a golden
detection channel in the recent experimental discovery. We have
parameterized the electromagnetic interactions of the Higgs boson by
an effective Lagrangian, and relate it by a low-energy theorem in
standard model to the Heisenberg-Euler effective Lagrangian of
electromagnetism that is induced by vacuum charged particles. For
the two-photon decay width, we recover the known result for the
leading contribution that accounts for about $65\%$ of the exact
one-loop result. The four- and six-photon decays are found to be
very much suppressed by phase space, in addition to naive counts of
the fine-structure constant.

\vspace{0.5cm}
\noindent %
{\bf Acknowledgement}

I would like to thank V.V. Skalozub for providing a copy of his
papers \cite{Skalozub:1975ab,Skalozub:1980nv,Skalozub:1986he} and L.
Ren for helping find a copy of the papers
\cite{Vanyashin,Skalozub:1975ab,Skalozub:1980nv,Skalozub:1986he}.
This work was supported in part by the grant NSFC-11025525 and by
the Fundamental Research Funds for the Central Universities
No.65030021.

\vspace{0.5cm}
\noindent %


\begin{thebibliography}{100}

\bibitem{:2012gk}
  G.~Aad {\it et al.}  [ATLAS Collaboration],
  Phys.\ Lett.\ B {\bf 716}, 1 (2012)  [arXiv:1207.7214 [hep-ex]].

\bibitem{:2012gu}
  S.~Chatrchyan {\it et al.}  [CMS Collaboration],
  Phys.\ Lett.\ B {\bf 716}, 30 (2012)  [arXiv:1207.7235 [hep-ex]].

\bibitem{Weinberg:1967tq}
  S.~Weinberg,
  Phys.\ Rev.\ Lett.\  {\bf 19}, 1264 (1967).

\bibitem{Glashow:1961tr}
  S.~L.~Glashow,
  Nucl.\ Phys.\  {\bf 22}, 579 (1961).

\bibitem{Salam:1968rm}
  A.~Salam,
  Conf.\ Proc.\ C {\bf 680519}, 367 (1968).

\bibitem{Gunion:1989we}
  J.~F.~Gunion, H.~E.~Haber, G.~L.~Kane and S.~Dawson,
  Front.\ Phys.\  {\bf 80}, 1 (2000);
  Errata: hep-ph/9302272.

\bibitem{Higgs:1964pj}
  P.~W.~Higgs,
  Phys.\ Rev.\ Lett.\  {\bf 13}, 508 (1964).

\bibitem{Englert:1964et}
  F.~Englert and R.~Brout,
  Phys.\ Rev.\ Lett.\  {\bf 13}, 321 (1964).

\bibitem{Guralnik:1964eu}
  G.~S.~Guralnik, C.~R.~Hagen and T.~W.~B.~Kibble,
  Phys.\ Rev.\ Lett.\  {\bf 13}, 585 (1964).

\bibitem{Ellis:1975ap}
  J.~R.~Ellis, M.~K.~Gaillard and D.~V.~Nanopoulos,
  Nucl.\ Phys.\ B {\bf 106}, 292 (1976).

\bibitem{Ioffe:1976sd}
  B.~L.~Ioffe and V.~A.~Khoze,
  Sov.\ J.\ Part.\ Nucl.\ {\bf 9}, 50 (1978) [Fiz.\ Elem.\ Chast.\ Atom.\ Yadra
  {\bf 9}, 118 (1978)].

\bibitem{Shifman:1979eb}
  M.~A.~Shifman, A.~I.~Vainshtein, M.~B.~Voloshin and V.~I.~Zakharov,
  Sov.\ J.\ Nucl.\ Phys.\  {\bf 30}, 711 (1979)  [Yad.\ Fiz.\  {\bf 30}, 1368 (1979)].

\bibitem{Vainshtein:1980ea}
  A.~I.~Vainshtein, V.~I.~Zakharov and M.~A.~Shifman,
  Sov.\ Phys.\ Usp.\  {\bf 23}, 429 (1980)  [Usp.\ Fiz.\ Nauk {\bf 131}, 537 (1980)].

\bibitem{Rizzo:1979mf}
  T.~G.~Rizzo,
  Phys.\ Rev.\ D {\bf 22}, 178 (1980)  [Addendum-ibid.\ D {\bf 22}, 1824 (1980)].

\bibitem{Karplus:1950zz}
  R.~Karplus and M.~Neuman,
  Phys.\ Rev.\  {\bf 83}, 776 (1951).

\bibitem{Aghababaie:2000zi}
  Y.~Aghababaie and C.~P.~Burgess,
  Phys.\ Rev.\ D {\bf 63}, 113006 (2001)  [hep-ph/0006165].

\bibitem{Heisenberg:1935qt}
  W.~Heisenberg and H.~Euler,
  Z.\ Phys.\  {\bf 98}, 714 (1936)  [physics/0605038].

\bibitem{Schwinger:1951nm}
  J.~S.~Schwinger,
  Phys.\ Rev.\  {\bf 82}, 664 (1951).

\bibitem{Dunne:2004nc}
  G.~V.~Dunne,
  In {\em From fields to strings, vol. 1}, editted by M. Shifman et al., pp.445-522
  [hep-th/0406216].

\bibitem{Vanyashin}
  V.S. Vanyashin and M.V. Terent'ev,
  Sov.\ Phys.\ JETP\ {\bf 21}, 375 (1965)
  [J.\ Exptl.\ Theoret.\ Phys.\ (USSR) {\bf 48}, 565 (1965)].

\bibitem{Skalozub:1975ab}
  V.~V. Skalozub,
  Sov.\ J.\ Nucl.\ Phys.\ {\bf 21}, 690 (1976)
  [Yad.\ Fiz.\  {\bf 21}, 1337 (1975)].

\bibitem{Skalozub:1980nv}
  V.~V.~Skalozub,
  Sov.\ J.\ Nucl.\ Phys.\ {\bf 31}, 412 (1980) [Yad.\ Fiz.\  {\bf 31}, 798 (1980)].

\bibitem{Skalozub:1986he}
  V.~V.~Skalozub,
  Sov.\ J.\ Part.\ Nucl.\ {\bf 16}, 445 (1985)
  [Fiz.\ Elem.\ Chast.\ Atom.\ Yadra {\bf 16}, 1005 (1985)].





\end{thebibliography}
\end{document}